\documentclass[aps,twocolumn,groupedaddress]{revtex4}
\usepackage{bm}
\usepackage{amsmath}

\begin{document}
\bibliographystyle{apsrev}

\title{External-field shifts of the $^{199}$Hg$^+$ optical frequency
standard\cite{thanks}}
\author{Wayne M. Itano}
\affiliation{Time and Frequency Division,
National Institute of Standards and Technology,
Boulder, CO 80305}

\date{12 March 2001}

\begin{abstract}
Frequency shifts
of the $^{199}$Hg$^+$ $5d^{10}6s$ $^2\mathrm{S}_{1/2}$ $(F=0, M_F=0)$
to $5d^9 6s^2$ $^2\mathrm{D}_{5/2}$ $(F=2, M_F=0)$ electric-quadrupole transition
at 282 nm due to external fields are calculated, based on a combination of measured
atomic parameters and {\em ab initio} calculations.
This transition is under investigation as an optical frequency standard.
The perturbations calculated are the quadratic Zeeman shift, the scalar and
tensor quadratic Stark shifts, and the interaction between an external
electric field gradient and the atomic quadrupole moment.
The quadrupole shift is likely to be the most difficult to evaluate
in a frequency standard and may have a magnitude of about 1 Hz for a
single ion in a Paul trap.

\end{abstract}
\maketitle

\section*{1. Introduction}

It has long been recognized that a frequency standard could be based on the
282 nm transition between the ground $5d^{10}6s$ $^2\mathrm{S}_{1/2}$ level and the
metastable $5d^9 6s^2$ $^2\mathrm{D}_{5/2}$ level of Hg$^+$ \cite{bender76}.
The lifetime of the upper level is $86(3)$ ms \cite{itano87},
so the ratio of the natural linewidth $\Delta\nu$ to the transition
frequency $\nu_0$ is $2\times 10^{-15}$.
(Unless otherwise noted, all uncertainties given in this paper are standard
uncertainties, i.e., one standard deviation estimates.)
Doppler broadening can be avoided if the transition is excited with two
counter-propagating photons, as originally proposed by Bender {\em et al.}
\cite{bender76} and subsequently demonstrated by Bergquist {\em et al.}
\cite{bergquist85}.  However, optical Stark
shifts are greatly reduced if the transition is driven instead with a single
photon by the electric-quadrupole interaction.  In this case, Doppler broadening
can be eliminated if the ion is confined to dimensions much
less than the optical wavelength, as was first demonstrated by
Bergquist {\em et al.} \cite{bergquist87}.

Recently, the $(F=0, M_F=0)$ to $(F=2, M_F=0)$ hyperfine component of the
$^{199}$Hg$^+$ $5d^{10}6s$ $^2\mathrm{S}_{1/2}$ to $5d^9 6s^2$ $^2\mathrm{D}_{5/2}$
single-photon transition has been
observed with a linewidth of only 6.7 Hz by Rafac {\em et al.} \cite{rafac00}.
A laser servo-locked to this transition is an extremely stable and
reproducible frequency reference.
New developments in optical frequency metrology \cite{reichert00,diddams00}
may soon make this system practical as an atomic frequency standard or clock.

While the $(F=0, M_F=0)$ to $(F=2, M_F=0)$ hyperfine component has no linear
Zeeman shift, it does have a quadratic Zeeman shift that must be accounted for.
In addition, there is a second-order Stark shift and a shift due to the
interaction between the electric-field gradient and the atomic electric-quadrupole
moment.
None of these shifts has yet been measured accurately, so it is useful to
have calculated values, even if they are not very precise.
Also, it is useful to know the functional form of the perturbation, even
if the magnitude is uncertain.
For example, the quadrupole shift can be eliminated by averaging the transition
frequency over three mutually orthogonal magnetic-field orientations, independent
of the orientation of the electric-field gradient.

\section*{2. Methods and notation}

The quadratic Zeeman shift can be calculated if the
hyperfine constants and electronic and nuclear $g$-factors are known.
Similarly, the quadratic Stark effect can be calculated from a knowledge
of the electric-dipole oscillator strengths.
The quadrupole shift depends on the atomic wavefunctions.
Some of these parameters have been measured, such as the hyperfine constants
and some of the oscillator strengths.
There are also published calculations for some of the oscillator strengths.

Here, we estimate, by the use of the Cowan atomic-structure codes,
values for parameters for which there are neither measured values nor
published calculations.
The Cowan codes are based on the Hartree-Fock approximation with some
relativistic corrections \cite{cowan81}.
The odd-parity configurations included in the calculation
were $5d^{10}np$ $(n=6,7,8,9)$, $5d^{10}5f$, $5d^96s6p$, $5d^96s7p$,
$5d^96s5f$, and $5d^86s^26p$.  The even-parity configurations were
$5d^{10}ns$ $(n=6,7,8,9,10)$, $5d^{10}nd$ $(n=6,7,8,9)$, $5d^96s^2$, $5d^96s7s$,
$5d^96s6d$, and $5d^9 6p^2$.
Recently, Sansonetti and Reader have made new measurements of the
spectrum of Hg$^+$ and classified many new lines \cite{sansonetti00}.
They also carried out a least-squares adjustment of the energy parameters
that enter the Cowan-code calculations in order to match the observed
energy levels.
We use these adjusted parameters in our Cowan-code calculations.

As one test of this method of calculation, we estimated the weakly
allowed 10.7 $\mu$m $5d^{10}6p$ $^2\mathrm{P}_{1/2}$ to $5d^96s^2$
$^2\mathrm{D}_{3/2}$ electric-dipole decay rate. This decay is
allowed only because of configuration mixing, since it requires
two electrons to change orbitals. The calculation shows the decay
to be due mostly to mixing between the $5d^{10}6p$ and $5d^9 6s
6p$ configurations. The calculated rate is 111 s$^{-1}$; the
measured rate is 52(16) s$^{-1}$ \cite{itano87}. Another test is
the electric-quadrupole decay rate of the $5d^96s^2$
$^2\mathrm{D}_{5/2}$ level to the ground level. The calculated
rate is 12.6 s$^{-1}$, and the measured rate is 11.6(0.4)
s$^{-1}$. Similar calculations have been carried out by Wilson
\cite{wilson90}.

Let $H_0$ be the atomic Hamiltonian, exclusive of the hyperfine and external
field effects, which are treated as perturbations.
For convenience, we denote the eigenstates of $H_0$ corresponding to the
electronic levels $5d^{10}6s$ $^2\mathrm{S}_{1/2}$
and $5d^9 6s^2$ $^2\mathrm{D}_{5/2}$ having $J_z$ eigenvalue $M_J$
by $|\mathrm{S} \;1/2\;M_J\rangle$ and $|\mathrm{D}\; 5/2\;M_J \rangle$,
respectively.

The corresponding eigenvalues of $H_0$ are denoted $W(\mathrm{S},1/2)$ and
$W(\mathrm{D},5/2)$.
An arbitrary eigenstate of $H_0$ with eigenvalue $W(\gamma,J)$ and electronic
angular momentum $J$ is denoted $|\gamma\; J\;M_J\rangle$.
Since $^{199}$Hg$^+$ has in addition a nuclear angular momentum $\bm{I}$,
where $I=1/2$, the complete state designation is $|\gamma J F M_F\rangle$,
where $F$ is the total angular momentum, and $M_F$ is the eigenvalue of $F_z$.

\section*{3. Quadratic Zeeman shift}

In order to calculate the energy shifts due to the hyperfine interaction
and to an external magnetic field $\bm{B}\equiv B\hat{\bm{z}}$,
we define effective Hamiltonian operators
$H^\prime_\mathrm{S}$ and $H^\prime_\mathrm{D}$ that operate within the subspaces
of hyperfine sublevels associated with the electronic levels
$5d^{10}6s$ $^2\mathrm{S}_{1/2}$ and $5d^9 6s^2$ $^2\mathrm{D}_{5/2}$ respectively:
\begin{equation}
H^\prime_\mathrm{S}=hA_\mathrm{S}\bm{I}\cdot\bm{J} +
g_J(\mathrm{S})\mu_{\mathrm{B}}\bm{J}\cdot\bm{B}+
g^\prime_I\mu_{\mathrm{B}}\bm{I}\cdot\bm{B},
\label{h_s}
\end{equation}
\begin{equation}
H^\prime_\mathrm{D}=hA_\mathrm{D}\bm{I}\cdot\bm{J} +
g_J(\mathrm{D})\mu_{\mathrm{B}}\bm{J}\cdot\bm{B}+
g^\prime_I\mu_{\mathrm{B}}\bm{I}\cdot\bm{B},
\label{h_d}
\end{equation}
where $A_\mathrm{S}$ and $A_\mathrm{D}$ are the dipole hyperfine constants,
$g_J(\mathrm{S})$ and $g_J(\mathrm{D})$ are the electronic $g$-factors,
$g^\prime_I$ is the nuclear $g$-factor, $h$ is the Planck constant,
and $\mu_{\mathrm{B}}$ is the Bohr magneton.
All of the parameters entering $H^\prime_\mathrm{S}$ and $H^\prime_\mathrm{D}$
are known from experiments, although a more accurate measurement of
$g_J(\mathrm{D})$ would be useful.
The ground-state hyperfine constant $A_\mathrm{S}$ has been measured in a
$^{199}$Hg$^+$ microwave frequency standard to be 40 507.347 996 841 59 (43) MHz
\cite{berkeland98}.
The excited-state hyperfine constant $A_\mathrm{D}$ has been measured recently
by an extension to the work described in Ref.\ \cite{rafac00},
in which the difference in the frequencies of the $|\mathrm{S}\;1/2\;0\;0\rangle$
to $|\mathrm{D}\;5/2\;2\;0\rangle$ and the $|\mathrm{S}\;1/2\;0\;0\rangle$
to $|\mathrm{D}\;5/2\;3\;0\rangle$ transition frequencies
was determined to be 3$A_\mathrm{D}$=2 958.57(12) MHz \cite{bergquist00},
in good agreement with an earlier, less precise measurement by
Fabry-P\'erot spectroscopy \cite{loebich62}.
The ground-state electronic $g$-factor $g_J(\mathrm{S})$ was measured in
$^{198}$Hg$^+$ by rf-optical double resonance to be 2.003 174 5(74) \cite{itano85}.
The excited-state electronic $g$-factor
$g_J(\mathrm{D})$ was measured in $^{198}$Hg$^+$ by conventional grating
spectroscopy of the 398 nm $5d^{10}6p$ $^2\mathrm{P}_{3/2}$ to $5d^96s^2$
$^2\mathrm{D}_{5/2}$ line to be 1.198 0(7) \cite{vankleef63}.
The difference in $g_J(\mathrm{S})$ or $g_J(\mathrm{D})$
between $^{198}$Hg$^+$ and $^{199}$Hg$^+$ is estimated to be much less
than the experimental uncertainties.
The nuclear $g$-factor $g^\prime_I$ is $-5.422\; 967(9) \times 10^{-4}$
\cite{cagnac61}.
The measurement was made with neutral ground-state $^{199}$Hg atoms, so the
diamagnetic shielding factor will be slightly different from that in the ion.
However, this is effect is negligible, since the magnitude of $g^\prime_I$
is so small compared to $g_J(\mathrm{S})$ or $g_J(\mathrm{D})$.

The determination of $g_J(\mathrm{D})$ could be improved by measuring the
optical-frequency difference
between two components of the 282 nm line and the frequency of a ground-state
microwave transition at the same magnetic field.
Since the uncertainty in the quadratic Zeeman shift is due mainly to the
uncertainty in $g_J(\mathrm{D})$, it is useful to see how accurately it can
be estimated theoretically.
The Land\'e $g$-factor for a $^2\mathrm{D}_{5/2}$ state, including the
correction for the anomalous magnetic moment of the electron, is 1.200 464.
The Cowan-code calculation shows that the configuration mixing does not
change this value by more than about $10^{-6}$, i.e., 1 in the last place.
There are several relativistic and diamagnetic corrections that
modify $g_J(\mathrm{D})$, one of which, called the Breit-Margenau correction
by Abragam and Van Vleck \cite{abragam53}, is proportional to the electron mean
kinetic energy.
The other corrections are more difficult to calculate.
The Cowan-code result for the mean kinetic energy of an electron in the
$5d$ orbital of the $5d^96s^2$ configuration is
$T=19.32\;hcR_\infty$, where $R_\infty$ is the Rydberg constant.
Using this value, we obtain a theoretical value of $g_J(\mathrm{D})$, including
the Breit-Margenau correction, of 1.199 85, which disagrees with the
the experimental value by $1.85\times 10^{-3}$, which is 2.6 times the
estimated experimental uncertainty of Ref.\ \cite{vankleef63}.
If we calculate $g_J(\mathrm{D})$ for neutral gold, which is isoelectronic
to Hg$^+$, by the same method, we obtain a value which differs
from the accurately measured experimental one \cite{childs66}
by $(7\pm 2)\times 10^{-5}$.
Thus, the error in the calculated value for $g_J(\mathrm{D})$ of
$^{199}$Hg$^+$ might be less than $1\times 10^{-4}$, but it is impossible to
be certain of this, since there are uncalculated terms.
Measurements of the $^{199}$Hg$^+$ optical clock frequency at different values
of the magnetic field
should result in a better experimental value for $g_J(\mathrm{D})$ in the near
future.

For low magnetic fields ($B$ less than 1 mT),
it is sufficient to calculate the energy levels to
second order in $B$.  To this order in $B$, the energies of the hyperfine-Zeeman
sublevels for the ground electronic level are
\begin{widetext}
\begin{eqnarray}
W(\mathrm{S},1/2,0,0,B)&=&W(\mathrm{S},1/2)-\frac{3hA_{\mathrm{S}}}{4}-
\frac{[g_J(\mathrm{S})-g^\prime_I]^2\mu_{\mathrm{B}}^2B^2}{4hA_{\mathrm{S}}},
\label{ground_00}\\
W(\mathrm{S},1/2,1,0,B)&=&W(\mathrm{S},1/2)+\frac{hA_{\mathrm{S}}}{4}+
\frac{[g_J(\mathrm{S})-g^\prime_I]^2\mu_{\mathrm{B}}^2B^2}{4hA_{\mathrm{S}}},
\label{ground_10}\\
W(\mathrm{S},1/2,1,\pm1,B)&=&W(\mathrm{S},1/2)+
\frac{hA_{\mathrm{S}}}{4}\pm\frac{[g_J(\mathrm{S})+
g^\prime_I]\mu_{\mathrm{B}} B}{2}.
\label{ground_1pm1}
\end{eqnarray}
For the $5d^96s^2$ $^2\mathrm{D}_{5/2}$ level we have
\begin{eqnarray}
W(\mathrm{D},5/2,2,0,B)&=&W(\mathrm{D},5/2)-\frac{7hA_{\mathrm{D}}}{4}-
\frac{[g_J(\mathrm{D})-g^\prime_I]^2\mu_{\mathrm{B}}^2B^2}{12hA_{\mathrm{D}}},
\label{dstate_20}\\
W(\mathrm{D},5/2,2,\pm 1,B)&=&W(\mathrm{D},5/2)-
\frac{7hA_{\mathrm{D}}}{4}\pm\frac{[7g_J(\mathrm{D})-
g^\prime_I]\mu_{\mathrm{B}} B}{6}
-\frac{2[g_J(\mathrm{D})-g^\prime_I]^2\mu_{\mathrm{B}}^2B^2}
{27hA_{\mathrm{D}}},\label{dstate_2pm1}\\
W(\mathrm{D},5/2,2,\pm 2,B)&=&W(\mathrm{D},5/2)-
\frac{7hA_{\mathrm{D}}}{4}\pm\frac{[7g_J(\mathrm{D})-
g^\prime_I]\mu_{\mathrm{B}} B}{3}
-\frac{5[g_J(\mathrm{D})-g^\prime_I]^2\mu_{\mathrm{B}}^2B^2}
{108hA_{\mathrm{D}}},\label{dstate_2pm2}\\
W(\mathrm{D},5/2,3,0,B)&=&W(\mathrm{D},5/2)+
\frac{5hA_{\mathrm{D}}}{4}+\frac{[g_J(\mathrm{D})-
g^\prime_I]^2\mu_{\mathrm{B}}^2B^2}{12hA_{\mathrm{D}}},
\label{dstate_30}\\
W(\mathrm{D},5/2,3,\pm 1,B)&=&W(\mathrm{D},5/2)+
\frac{5hA_{\mathrm{D}}}{4}\pm\frac{[5g_J(\mathrm{D})+
g^\prime_I]\mu_{\mathrm{B}} B}{6}
+\frac{2[g_J(\mathrm{D})-g^\prime_I]^2\mu_{\mathrm{B}}^2B^2}
{27hA_{\mathrm{D}}},\label{dstate_3pm1}\\
W(\mathrm{D},5/2,3,\pm 2,B)&=&W(\mathrm{D},5/2)+
\frac{5hA_{\mathrm{D}}}{4}\pm\frac{[5g_J(\mathrm{D})+
g^\prime_I]\mu_{\mathrm{B}} B}{3}\nonumber
+\frac{5[g_J(\mathrm{D})-g^\prime_I]^2\mu_{\mathrm{B}}^2B^2}
{108hA_{\mathrm{D}}},\label{dstate_3pm2}\\
W(\mathrm{D},5/2,3,\pm 3,B)&=&W(\mathrm{D},5/2)+
\frac{5hA_{\mathrm{D}}}{4}\pm\frac{[5g_J(\mathrm{D})+g^\prime_I]\mu_{\mathrm{B}}B}{2}.
\label{dstate_3pm3}
\end{eqnarray}
\end{widetext}
Here, $W(\gamma,J,F,M_F,B)$ denotes the energy of the state
$|\gamma J F M_F\rangle$, including the effects of the hyperfine interaction
and the magnetic field.

At a value of $B$ of 0.1 mT, the quadratic shift of the
$|\mathrm{S}\; 1/2\; 0\; 0\rangle$ to $|\mathrm{D}\; 5/2\; 2\; 0\rangle$
transition (optical clock transition)is $-189.25(28)$ Hz,
where the uncertainty stems mainly from the uncertainty in the
experimental value of $g_J(\mathrm{D})$.
In practice, the error may be less than this if the magnetic field is determined
from the Zeeman splittings within the $|\mathrm{D}\; 5/2\; F\; M_F\rangle$
sublevels.
The reason is that an error in $g_J(\mathrm{D})$ leads to an error in the
value of $B$ inferred from the Zeeman splittings, which partly compensates for
the $g_J(\mathrm{D})$ error.
If instead we use the calculated value of $g_J(\mathrm{D})$, the quadratic shift
for $B=0.1$ mT is $-189.98$ Hz, where the uncertainty is difficult to estimate.

\section*{4. Quadratic Stark shift}

The theory of the quadratic Stark shift in free atoms has been described
in detail by Angel and Sandars \cite{angel68}.
The Stark Hamiltonian is
\begin{equation}
H_E=-\bm{\mu}\cdot\bm{E},
\label{starkhamiltonian}
\end{equation}
where $\bm{\mu}$ is the electric-dipole moment operator,
\begin{equation}
\bm{\mu}=-e\sum_{i}\bm{r}_i,
\label{dipoleoperator}
\end{equation}
and $\bm{E}$ is the applied external electric field.
In Eq.~(\ref{dipoleoperator}), $\bm{r}_i$ is the position operator of the
$i$th electron, measured relative to the nucleus, and the summation is
over all electrons.

First consider an atom with zero nuclear spin, such as $^{198}$Hg$^+$.
To second order in the electric field, the Stark shifts of the
set of sublevels $|\gamma J M_J\rangle$ depend on two parameters,
$\alpha_{\text{scalar}}(\gamma,J)$ and $\alpha_{\text{tensor}}(\gamma,J)$,
called the scalar and tensor polarizabilities.
In principle, when both magnetic and electric fields
are present but are not parallel, the energy levels are obtained by
simultaneously diagonalizing
the hyperfine, Zeeman, and Stark Hamiltonians.
In practice, the Zeeman shifts are normally much larger than the Stark shifts,
so that $H_E$ does not affect the diagonalization.
In that case, the energy shift of the state $|\gamma J M_J\rangle$ due to $H_E$ is
\begin{widetext}
\begin{equation}
\Delta W_E(\gamma,J,M_J,\bm{E})=-\tfrac{1}{2}\alpha_{\text{scalar}}(\gamma,J)E^2
-\tfrac{1}{4}\alpha_{\text{tensor}}(\gamma,J)\frac{[3M^2_J-J(J+1)]}{J(2J-1)}(3E^2_z-E^2).
\label{starkshift1}
\end{equation}

Treating $H_E$ by second-order perturbation theory leads to the following
expressions for the polarizabilities \cite{angel68}:
\begin{eqnarray}
\alpha_{\text{scalar}}{(\gamma,J)}&=&
\frac{8\pi\epsilon_0}{3(2J+1)}\sum_{\gamma^\prime J^\prime}
\frac{|(\gamma J\|\mu^{(1)}\|\gamma^\prime J^\prime)|^2}
{W(\gamma^\prime, J^\prime)-W(\gamma,J)},
\label{alphascalar}\\
\alpha_{\text{tensor}}(\gamma,J)&=&8\pi\epsilon_0
\left[\frac{10J(2J-1)}{3(2J+3)(J+1)(2J+1)}\right]^{1/2}
\sum_{\gamma^\prime J^\prime}(-1)^{J-J^\prime}
\left\{ \begin{array}{ccc} 1&1&2\\J&J&J^\prime
\end{array}  \right\}
\frac{|(\gamma J\|\mu^{(1)}\|\gamma^\prime J^\prime)|^2}
{W(\gamma^\prime, J^\prime)-W(\gamma,J)}.
\label{alphatensor}
\end{eqnarray}
The summations are over all levels other than $|\gamma J\rangle$.
Equations (\ref{alphascalar}) and(\ref{alphatensor}) can be rewritten in terms
of the oscillator strengths $f_{\gamma J, \gamma^\prime J^\prime}$:
\begin{eqnarray}
\alpha_{\text{scalar}}{(\gamma,J)}&=&
\frac{4\pi\epsilon_0 e^2\hbar^2}{m_e}\sum_{\gamma^\prime J^\prime}
\frac{f_{\gamma J, \gamma^\prime J^\prime}}
{[W(\gamma^\prime, J^\prime)-W(\gamma,J)]^2},
\label{alphascalar2}\\
\alpha_{\text{tensor}}(\gamma,J)&=&\frac{4\pi\epsilon_0 e^2\hbar^2}{m_e}
\left[\frac{30J(2J-1)(2J+1)}{(2J+3)(J+1)}\right]^{1/2}
\sum_{\gamma^\prime J^\prime}(-1)^{J-J^\prime}
\left\{ \begin{array}{ccc} 1&1&2\\J&J&J^\prime
\end{array}  \right\}
\frac{f_{\gamma J, \gamma^\prime J^\prime}}{[W(\gamma^\prime, J^\prime)-W(\gamma,J)]^2},
\label{alphatensor2}
\end{eqnarray}
where $m_e$ is the electron mass.
The tensor polarizability is zero for levels with $J<1$, such as
the Hg$^+$ $5d^{10}6s$ $^2\mathrm{S}_{1/2}$ level.

For an atom with nonzero nuclear spin $I$, the quadratic Stark shift of the
state $|\gamma J F M_F\rangle$ is
\begin{equation}
\Delta W_E(\gamma,J,F,M_F,\bm{E})=-\tfrac{1}{2}\alpha_{\text{scalar}}(\gamma,J,F)E^2
-\tfrac{1}{4}\alpha_{\text{tensor}}(\gamma,J,F)\frac{[3M^2_F-F(F+1)]}{F(2F-1)}(3E^2_z-E^2).
\label{starkshift2}
\end{equation}
\end{widetext}

We make the approximation that hyperfine interaction does not modify the electronic
part of the atomic wavefunctions (the $IJ$-coupling approximation of Angel
and Sandars \cite{angel68}).
This approximation is adequate for the present purpose, which is to evaluate
the Stark shift of the $^{199}$Hg$^+$ optical clock transition.
Obtaining the differential Stark shift between the hyperfine levels
of the ground state, which is significant for the $^{199}$Hg$^+$ microwave
frequency standard \cite{berkeland98}, requires going to a higher order
of perturbation theory \cite{itano82}.
In the $IJ$-coupling approximation \cite{angel68},
\begin{widetext}
\begin{eqnarray}
\alpha_{\text{scalar}}(\gamma,J,F)&=&\alpha_{\text{scalar}}(\gamma,J),\label{scalarF}\\
\alpha_{\text{tensor}}(\gamma,J,F)&=&(-1)^{I+J+F}
\left[\frac{F(2F-1)(2F+1)(2J+3)(2J+1)(J+1)}{(2F+3)(F+1)J(2J-1)}\right]^{1/2}
\left\{ \begin{array}{ccc} F&J&I\\J&F&2\end{array}  \right\}
\alpha_{\text{tensor}}(\gamma,J).
\label{tensorF}
\end{eqnarray}
\end{widetext}

Equations (\ref{alphascalar2}) and (\ref{alphatensor2}) were used to evaluate
the polarizabilities for the Hg$^+$ $5d^{10}6s$ $^2\mathrm{S}_{1/2}$ and
$5d^9 6s^2$ $^2\mathrm{D}_{5/2}$ levels.
For the calculation of $\alpha_{\text{scalar}}(\mathrm{S},1/2)$,
the oscillator strengths for all electric-dipole transitions connecting the
$5d^{10}6s$ configuration to the
$5d^{10}np$ $(n=6,7,8)$ and $5d^9 6s6p$ configurations were included.
These were taken from the theoretical work of Brage {\em et al.} \cite{brage99}.
The final result is
$\alpha_{\text{scalar}}(\mathrm{S},1/2)/(4\pi\epsilon_0)=2.41\times 10^{-24}$
cm$^3$, which compares very well with the value of $2.22\times 10^{-24}$ cm$^3$
obtained by Henderson {\em et al.} from a combination of experimental and
calculated oscillator strengths \cite{henderson97}.
For the calculations of $\alpha_{\text{scalar}}(\mathrm{D},5/2)$ and
$\alpha_{\text{tensor}}(\mathrm{D},5/2)$, the oscillator strengths for
electric-dipole transitions to the $5d^{10}np$ $(n=6,7,8)$, $5d^{10}5f$,
and $5d^9 6s6p$ configurations were taken from Brage {\em et al.} \cite{brage99}.
The oscillator strengths for electric-dipole transitions to the $5d^9 6s7p$
and $5d^8 6s^2 6p$ configurations were taken from the Cowan-code calculations.
The results were $\alpha_{\text{scalar}}(\mathrm{D},5/2)/(4\pi\epsilon_0)
=3.77\times 10^{-24}$ cm$^3$ and
$\alpha_{\text{tensor}}(\mathrm{D},5/2)/(4\pi\epsilon_0)=-0.263\times 10^{-24}$
cm$^3$.
Evaluating Eq.~(\ref{tensorF}) for $F$=2 and $F$=3 in the
$5d^96s^2$ $^2\mathrm{D}_{5/2}$ level,
we obtain $\alpha_{\text{tensor}}(\mathrm{D},5/2,2)
=\tfrac{4}{5}\alpha_{\text{tensor}}(\mathrm{D},5/2)$
and $\alpha_{\text{tensor}}(\mathrm{D},5/2,3)=
\alpha_{\text{tensor}}(\mathrm{D},5/2)$.

The tensor polarizability is much smaller than the scalar polarizabilities and
in any case does not contribute if the external electric field is isotropic, as
is the case for the blackbody radiation field.
The net shift of the optical clock transition due to the scalar polarizabilities is
$\tfrac{1}{2}[\alpha_{\text{scalar}}(\mathrm{S},1/2)-\alpha_{\text{scalar}}
(\mathrm{D},5/2)]E^2$.
In frequency units, the shift is $-1.14\times 10^{-3}$ $E^2$ Hz, where $E$ is
expressed in V/cm.
The error in the coefficient is difficult to estimate, particularly since it
is a difference of two quantities of about the same size.
However, the total shifts are small for typical experimental conditions.
If the electric field is time-dependent, as for the blackbody field, the mean-squared
value $\langle E^2\rangle$ is taken.
At a temperature of 300 K, the shift of the optical  clock transition
due to the blackbody electric field is $-0.079$ Hz.
The mean-squared blackbody field is proportional to the fourth power of the
temperature.
For a single, laser-cooled ion in a Paul trap, the mean-squared trapping
electric fields can be made small enough that the Stark shifts are not likely to
be observable \cite{berkeland98a}.

\section*{5. Electric quadrupole shift}

The atomic quadrupole moment is due to a departure of the electronic charge
distribution of an atom from spherical symmetry.
Atomic quadrupole moments were first measured by the shift in energy levels
due to an applied electric-field gradient in atomic-beam resonance experiments
\cite{angel67,sandars73}.

The interaction of the atomic quadrupole moment with external
electric-field gradients, for example those generated by the electrodes of an ion
trap, is analogous to the interaction of a nuclear quadrupole moment
with the electric field gradients due to the atomic electrons.
Hence, we can adapt the treatment used for the electric-quadrupole hyperfine
interaction of an atom \cite{ramsey56}.
The Hamiltonian describing the interaction of external electric-field gradients
with the atomic quadrupole moment is
\begin{equation}
H_\mathrm{Q}=\bm{\nabla E}^{(2)}\cdot \bm{\mathit{\mathit{\Theta}}}^{(2)}
=\sum_{q=-2}^{2}(-1)^q\nabla E^{(2)}_q \mathit{\mathit{\Theta}}^{(2)}_{-q},
\label{hquad}
\end{equation}
where $\bm{\nabla E}^{(2)}$ is a tensor describing the gradients of the external
electric field at the position of the atom, and $\bm{\mathit{\Theta}}^{(2)}$ is the
electric-quadrupole operator for the atom.

Following Ref.~\cite{ramsey56}, we define the components of
$\bm{\nabla E}^{(2)}$ as
\begin{eqnarray}
\nabla E^{(2)}_0&=&-\frac{1}{2}\;\frac{\partial E_z}{\partial z},\\
\nabla E^{(2)}_{\pm 1}&=&\pm \frac{\sqrt{6}}{6}\;\frac{\partial E_\pm}{\partial z}=
\pm\frac{\sqrt{6}}{6}\;\partial_\pm E_z,\\
\nabla E^{(2)}_{\pm 2}&=&-\frac{\sqrt{6}}{12}\;\partial_\pm E_\pm,
\end{eqnarray}
where $E_\pm\equiv E_x \pm \mathrm{i} E_y$ and
$\partial_\pm \equiv \frac{\partial}{\partial x}\pm
\mathrm{i}\frac{\partial}{\partial y}$.

The operator components $\mathit{\Theta}^{(2)}_q$ are defined in terms of the electronic
coordinate operators as
\begin{eqnarray}
\mathit{\Theta}^{(2)}_0&=&-\frac{e}{2}\sum_j (3z_j^2-r_j^2),\label{theta20}\\
\mathit{\Theta}^{(2)}_{\pm 1}&=&-e\;\sqrt{\frac{3}{2}}\sum_j z_j (x_j\pm \mathrm{i}y_j),\label{theta21}\\
\mathit{\Theta}^{(2)}_{\pm 2}&=&-e\;\sqrt{\frac{3}{8}}\sum_j (x_j\pm \mathrm{i}y_j)^2,\label{theta22}
\end{eqnarray}
where the sums are taken over all the electrons.
The quadrupole moment $\mathit{\Theta}(\gamma,J)$ of an atomic level $|\gamma J\rangle$
is defined by the diagonal matrix element in the state with maximum $M_J$:
\begin{equation}
\mathit{\Theta}(\gamma,J)=\langle \gamma J J|\mathit{\Theta}^{(2)}_0|\gamma J J\rangle.
\label{quadmoment}
\end{equation}
This is the definition used by Angel {\em et al.} \cite{angel67}.

In order to simplify the form of $\bm{\nabla E}^{(2)}$, we make a principal-axis
transformation as in Ref.~\cite{brown82}.
That is, we express the electric potential in the neighborhood of the atom as
\begin{equation}
\mathit{\Phi}(x^\prime,y^\prime,z^\prime)=A[({x^\prime}^2+{y^\prime}^2-2{z^\prime}^2)
+\epsilon({x^\prime}^2-{y^\prime}^2)].
\label{phi_simple}
\end{equation}
The principal-axis (primed) frame
($x^\prime$,$y^\prime$,$z^\prime$) is the one in which
$\mathit{\Phi}$ has the simple form of Eq.~(\ref{phi_simple}),
while the laboratory (unprimed)
frame ($x$,$y$,$z$) is the  in which the magnetic field is oriented along the $z$ axis.

The tensor components of $\bm{\nabla E}^{(2)}$ in the principal-axis frame
are obtained by taking derivatives of $\mathit{\Phi}(x^\prime,y^\prime,z^\prime)$:
\begin{eqnarray}
{\nabla E^{(2)}_0}^\prime&=&-2A,\label{e20}\\
{\nabla E^{(2)}_{\pm 1}}^\prime&=&0,\label{e21}\\
{\nabla E^{(2)}_{\pm 2}}^\prime&=&\sqrt{\frac{2}{3}}\;\epsilon A.\label{e22}
\end{eqnarray}
In the principal-axis frame, $H_\mathrm{Q}$ has the simple form
\begin{equation}
H_\mathrm{Q}=-2A{\mathit{\Theta}^{(2)}_0}^\prime+\sqrt{\frac{2}{3}}\;\epsilon A\left({\mathit{\Theta}^{(2)}_2}^\prime
+{\mathit{\Theta}^{(2)}_{-2}}^\prime\right).
\label{hquad-simple}
\end{equation}

As long as the energy shifts due to $H_\mathrm{Q}$ are small relative to the Zeeman shifts, which
is the usual case in practice, $H_\mathrm{Q}$ can be treated as a perturbation.
In that case, it is necessary only to evaluate the matrix elements of $H_\mathrm{Q}$
that are diagonal in the basis of states $|\gamma J F M_F\rangle$, where $\bm{F}$
is the total atomic angular momentum, including nuclear spin $\bm{I}$, and
$M_F$ is the eigenvalue of
$F_z$ with respect to the laboratory (not principal-axis) frame.
Let $\bm{\omega}$ denote the set of Euler angles $\{\alpha,\beta,\gamma\}$
that takes the principal-axis frame to the laboratory frame.
To be explicit, starting from the principal-axis frame, we rotate the
coordinate system about the $z$ axis by $\alpha$, then about the new $y$ axis by
$\beta$, and then about the new $z$ axis by $\gamma$ so that the rotated coordinate
system coincides with the laboratory coordinate system.
We can set $\gamma=0$, since the final rotation about the laboratory $z$ axis, which
is parallel to $\bm{B}$, has no effect.
The states $|\gamma J F m\rangle^\prime$ defined in the principal-axis frame and the
states $|\gamma J F \mu\rangle$ defined in the laboratory frame are related by
\begin{equation}
|\gamma J F m\rangle^\prime = \sum_{\mu}D^{(F)}_{\mu m}(\bm{\omega})
|\gamma J F \mu\rangle,
\label{transform}
\end{equation}
where $D^{(F)}_{\mu m}(\bm{\omega})$ is a rotation matrix element defined in
the passive representation \cite{edmonds74,wolf69}.
The inverse relation is
\begin{equation}
|\gamma J F \mu\rangle=\sum_{m}{D^{(F)}_{\mu m}}^*(\bm{\omega})
|\gamma J F m\rangle^\prime.
\label{inverse}
\end{equation}

In order to evaluate the diagonal matrix elements of $H_\mathrm{Q}$ in the laboratory
frame, it is necessary to evaluate matrix elements of the operators
${\mathit{\Theta}^{(2)}_q}^\prime$, defined in the principal-axis frame.
These matrix elements are of the form
\begin{widetext}
\begin{eqnarray}
\lefteqn{\langle \gamma J F \mu|{\mathit{\Theta}^{(2)}_q}^\prime|\gamma J F\mu\rangle=
\sum_{m^\prime\; m}D^{(F)}_{\mu m^\prime}(\bm{\omega})
{D^{(F)}_{\mu m}}^*(\bm{\omega})\;\;^\prime\langle\gamma J F m^\prime|{\mathit{\Theta}^{(2)}_q}^\prime
|\gamma J F m\rangle^\prime,}\label{quad1}\\
&=&(\gamma J F \|\mathit{\Theta}^{(2)}\|\gamma J F)\sum_{m^\prime\; m}(-1)^{F-m^\prime}
\left(\begin{array}{ccc} F&2&F\\-m^\prime&q&m\end{array}\right)
D^{(F)}_{\mu m^\prime}(\bm{\omega}){D^{(F)}_{\mu m}}^*(\bm{\omega}),
\label{quad2}\\
&=&(-1)^{F-\mu-q}(\gamma J F \|\mathit{\Theta}^{(2)}\|\gamma J F)\sum_{m^\prime\; m}
\left(\begin{array}{ccc} F&2&F\\-m^\prime&q&m\end{array}\right)
D^{(F)}_{\mu m^\prime}(\bm{\omega})D^{(F)}_{-\mu -m}(\bm{\omega}),
\label{quad3}\\
&=&(-1)^{F-\mu-q}(\gamma J F \|\mathit{\Theta}^{(2)}\|\gamma J F)\sum_{K\;m\;m^\prime\;n\;n^\prime}
(2K+1)\left(\begin{array}{ccc} F&2&F\\-m^\prime&q&m\end{array}\right)
\left(\begin{array}{ccc} F&F&K\\\mu&-\mu&n^\prime\end{array}\right)
\left(\begin{array}{ccc} F&F&K\\ m^\prime&-m&n\end{array}\right)
{D^{(K)}_{n^\prime n}}^*(\bm{\omega}),\label{quad4}\\
&=&(-1)^{F-\mu-q}(\gamma J F \|\mathit{\Theta}^{(2)}\|\gamma J F)
\left(\begin{array}{ccc} F&2&F\\ -\mu&0&\mu\end{array}\right)
{D^{(2)}_{0 -q}}^*(\bm{\omega}),\label{quad5}
\end{eqnarray}
where Eq.~(\ref{quad2}) follows from the Wigner-Eckart theorem, and Eqs.~(\ref{quad3}),
(\ref{quad4}), and (\ref{quad5}) follow from Eqs.~(4.2.7), (4.3.2), and (3.7.8)
of Ref.~\cite{edmonds74}, respectively.
The required rotation matrix elements are, from Eq.~(4.1.25) of Ref.~\cite{edmonds74}
(with correction of a typographical error),
\begin{eqnarray}
{D^{(2)}_{0 0}}^*(\bm{\omega})&=&\tfrac{1}{2}(3\cos^2 \beta -1),\\
{D^{(2)}_{0\; \pm 2}}^*(\bm{\omega})&=&\sqrt{\tfrac{3}{8}}\sin^2 \beta
(\cos 2\alpha \mp \mathrm{i}\sin 2\alpha).
\end{eqnarray}
The 3-$j$ symbol in Eq.~(\ref{quad5}) is
\begin{equation}
\left(\begin{array}{ccc} F&2&F\\ -\mu&0&\mu\end{array}\right)=(-1)^{F-\mu}
\frac{2[3\mu^2-F(F+1)]}{[(2F+3)(2F+2)(2F+1)2F(2F-1)]^{1/2}}.
\end{equation}
The diagonal matrix elements of $H_\mathrm{Q}$ in the laboratory frame are
\begin{eqnarray}
\lefteqn{\langle \gamma J F M_F|H_\mathrm{Q}|\gamma J F M_F\rangle=}\nonumber\\
& & \frac{-2[3M_F^2-F(F+1)]A(\gamma J F \|\mathit{\Theta}^{(2)}\|\gamma J F)}
{[(2F+3)(2F+2)(2F+1)2F(2F-1)]^{1/2}}
[3\cos^2\beta-1)-\epsilon\sin^2\beta(\cos^2\alpha-\sin^2\alpha)].
\label{generalhq}
\end{eqnarray}
\end{widetext}

It is simple to show, by directly integrating the angular factor
in square brackets in Eq.~(\ref{generalhq}), that the average value of the
diagonal matrix elements of $H_\mathrm{Q}$, taken over all
possible orientations of the laboratory frame with respect to the
principal-axis frame, is zero.
This also follows directly from the fact that the quantity in square
brackets is a linear combination of spherical harmonics.
It is less obvious that the average, taken over any three mutually perpendicular
orientations of the laboratory $z$ quantization axis, is also zero.
This result is proven in the Appendix.
This provides a method for eliminating the quadrupole shift from the
observed transition frequency.
The magnetic field must be oriented in three mutually perpendicular
directions with respect to the trap electrodes, which are the source of the
external quadrupole field, but with the same magnitude of the magnetic field.
The average of the transition frequencies taken under these three conditions does not
contain the quadrupole shift.

The reduced matrix element in Eq.~(\ref{generalhq}) is, in the $IJ$-coupling
approximation,
\begin{widetext}
\begin{equation}
(\gamma (IJ)F\|\mathit{\Theta}^{(2)}\|\gamma (IJ)F)=
(-1)^{I+J+F}(2F+1)\left\{\begin{array}{ccc} J&2&J\\ F&I&F\end{array}\right\}
\left(\begin{array}{ccc} J&2&J\\ -J&0&J\end{array}\right)^{-1}\mathit{\Theta}(\gamma,J),
\end{equation}
\end{widetext}
where $I$ is included in the state notation in order to specify the order of
coupling of $I$ and $J$.
For the particular case of the $^{199}$Hg$^+$ $5d^96s^2$ $^2\mathrm{D}_{5/2}$
level, the reduced matrix elements are
\begin{equation}
(\mathrm{D}\; 5/2 \;2\|\mathit{\Theta}^{(2)}\|\mathrm{D}\; 5/2\; 2)=
 2\;\sqrt{\frac{14}{5}}\;\mathit{\Theta}(\mathrm{D},5/2),
\end{equation}
\begin{equation}(\mathrm{D}\; 5/2 \;3\|\mathit{\Theta}^{(2)}\|\mathrm{D}\; 5/2\; 3)=
 2\;\sqrt{\frac{21}{5}}\;\mathit{\Theta}(\mathrm{D},5/2).
\end{equation}

Since the Cowan-code calculation shows that there is very little configuration mixing
in the $^{199}$Hg$^+$ $5d^96s^2$ $^2\mathrm{D}_{5/2}$ level,
$\mathit{\Theta}(\mathit{D},5/2)$ can be reduced
to a matrix element involving only the $5d$ orbital:
\begin{widetext}
\begin{eqnarray}
\mathit{\Theta}(\mathit{D},5/2)&=&\frac{e}{2}\;\langle 5d\, ^2d_{5/2},\;m_j=5/2|3z^2-r^2|5d\, ^2d_{5/2},\;m_j=5/2
\rangle,\label{quad5d}\\
&=&\frac{e}{2}\;\langle 5d,\;m_l=2|3z^2-r^2|5d,\;m_l=2\rangle,\\
&=&e\sqrt{\frac{4\pi}{5}}\langle 5d,\;m_l=2|Y_{2\,0}(\theta,\phi)|5d,\;m_l=2\rangle,\\
&=&e\sqrt{\frac{4\pi}{5}}\langle 5d| r^2|5d\rangle
\int_0^{2\pi}\int_0^\pi  Y_{2\,2}^*(\theta,\phi)
Y_{2\,0}(\theta,\phi)Y_{2\,2}(\theta,\phi)\sin\theta d\theta d\phi,\\
&=&5e\langle 5d| r^2|5d\rangle \left(\begin{array}{ccc} 2&2&2\\ -2&0&2\end{array}\right)
\left(\begin{array}{ccc} 2&2&2\\ 0&0&0\end{array}\right),\\
&=&-\frac{2e}{7}\langle 5d| r^2|5d\rangle.
\end{eqnarray}
The apparent sign reversal in Eq.~(\ref{quad5d}) relative to Eqs.~(\ref{theta20})
and (\ref{quadmoment})
is due to the fact that the quadrupole moment is due to a single {\em hole}
in the otherwise filled $5d$ shell rather than to a single {\em electron}.
According to the Cowan-code calculation,
\begin{equation}
\langle 5d|r^2|5d\rangle=2.324\; a_0^2=6.509\times 10^{-17}\; \mathrm{cm}^2,
\end{equation}
where $a_0$ is the Bohr radius.

Since the quadrupole shifts are zero in the $5d^{10}6s$ $^2\mathrm{S}_{1/2}$ level,
the quadrupole shift of the $^{199}$Hg$^+$ optical clock transition
is due entirely to the shift of the $|\mathrm{D}\;5/2\;2\;0\rangle$ state, and is
given by
\begin{eqnarray}
\langle \mathrm{D}\;5/2\;2\;0|H_\mathrm{Q}|\mathrm{D}\;5/2\;2\;0\rangle
&=&
\tfrac{4}{5}A\mathit{\Theta}(\mathrm{D},5/2)
[(3\cos^2\beta-1)-\epsilon\sin^2\beta(\cos^2\alpha-\sin^2\alpha)],\\
&=&-\tfrac{8}{35}Ae\langle 5d|r^2|5d\rangle
[(3\cos^2\beta-1)-\epsilon\sin^2\beta(\cos^2\alpha-\sin^2\alpha)],\\
&\approx& -3.6\times 10^{-3}hA
[(3\cos^2\beta-1)-\epsilon\sin^2\beta(\cos^2\alpha-\sin^2\alpha)]\; \mathrm{Hz},
\end{eqnarray}
where $A$ is expressed in units of V/cm$^2$.
Thus, for typical values $A\approx$ 10$^3$ V/cm$^2$ and $|\epsilon|\alt 1$,
the quadrupole shift is on the order of 1 Hz.

\begin{acknowledgments}
We thank Dr.\ C.\ J.\ Sansonetti for making available the results of
Ref.~\cite{sansonetti00} prior to publication.
We acknowledge financial support from the U.S. Office of Naval Research.
\end{acknowledgments}

\appendix

\section*{6. Appendix. Angular averaging of the quadrupole shift}

For the purpose of describing the quadrupole shift, the orientation of the
laboratory (quantization) axis with respect to the principal-axis frame
is defined by the angles $\beta$ and $\alpha$.
In the principal-axis coordinate system, a unit vector along the laboratory $z$ axis
is defined in terms of $\beta$ and $\alpha$ by
\begin{equation}
\hat{\bm{z}}=(\sin\beta \cos\alpha, \sin\beta \sin\alpha, \cos\beta).
\label{unitz}
\end{equation}
We wish to show that the angular dependence of the quadrupole shift is such that
the diagonal matrix elements given by Eq.~(\ref{generalhq}) average to zero,
for $\hat{\bm{z}}$ along any three mutually perpendicular directions.

An arbitrary set of three mutually perpendicular unit vectors $\bm{e}_1$,
$\bm{e}_2$, and $\bm{e}_3$ can be
parameterized by the set of angles $\theta$, $\phi$, and $\psi$ in the
following way:
\begin{eqnarray}
\bm{e}_1&=&(\sin\theta\cos\phi, \sin\theta\sin\phi, \cos\theta),\\
\bm{e}_2&=&(\cos\phi\cos\theta\cos\psi-\sin\phi\sin\psi,
\sin\phi\cos\theta\cos\psi+\cos\phi\sin\psi, -\sin\theta\cos\psi),\\
\bm{e}_3&=&(-\cos\phi\cos\theta\sin\psi-\sin\phi\cos\psi,
-\sin\phi\cos\theta\sin\psi+\cos\phi\cos\psi, \sin\theta\sin\psi).
\end{eqnarray}
It can be verified by direct computation that
$\bm{e}_i\cdot\bm{e}_j=\delta_{ij}$.

The quadrupole shift can be evaluated for each of these three unit vectors
substituted for $\hat{\bm{z}}$ [Eq.~(\ref{unitz})] and the average taken.
First consider the average of the quantity $(3\cos^2 \beta-1)$
that appears in Eq.~(\ref{generalhq}):
We use the fact that $\cos\beta$ is the third component of $\hat{\bm{z}}$,
so the average is:
\begin{eqnarray}
\langle 3\cos^2\beta -1 \rangle&=&\cos^2\theta + \sin^2\theta\cos^2\psi
+\sin^2\theta\sin^2\psi-1,\\
&=&\cos^2\theta+\sin^2\theta-1,\\
&=&0 ,
\end{eqnarray}
for arbitrary $\theta$, $\phi$, and $\psi$.
Similarly, the average of the other angle-dependent term in Eq.~(\ref{generalhq}),
$\sin^2\beta(\cos^2\alpha-\sin^2\alpha)$, is
calculated by making use of the fact that $\sin\beta\cos\alpha$ is the first
component of $\hat{\bm{z}}$, and $\sin\beta\sin\alpha$ is the second:
\begin{eqnarray}
\langle \sin^2\beta(\cos^2\alpha-\sin^2\alpha)\rangle
&=&
\tfrac{1}{3}[\sin^2\theta\cos^2\phi-\sin^2\theta\sin^2\phi\nonumber\\
& &+(\cos\phi\cos\theta\cos\psi-\sin\phi\sin\psi)^2-
(\sin\phi\cos\theta\cos\psi+\cos\phi\sin\psi)^2\nonumber\\
& &+(\cos\phi\cos\theta\sin\psi+\sin\phi\cos\psi)^2
-(\sin\phi\cos\theta\sin\psi-\cos\phi\cos\psi)^2],\\
&=&0,
\end{eqnarray}
\end{widetext}
for arbitrary $\theta$, $\phi$, and $\psi$.
Hence, the matrix elements of $H_\mathrm{Q}$ given by Eq.~(\ref{generalhq})
average to zero for any three mutually perpendicular orientations of the
laboratory quantization axis.

\section*{About the author:}
Wayne M. Itano is a physicist in the Time and Frequency Division, Physics
Laboratory, National Institute of Standards and Technology.
The National Institute of Standards and Technology is an agency of the Technology
Administration, U.S. Department of Commerce.
\end{document}